\documentclass[12pt]{JHEP3}
\usepackage{epsfig}

\def\fun#1#2{\lower3.6pt\vbox{\baselineskip0pt\lineskip.9pt
  \ialign{$\mathsurround=0pt#1\hfil##\hfil$\crcr#2\crcr\sim\crcr}}}
\newcommand{\beq}{\begin{equation}}
\newcommand{\eeq}{\end{equation}}
\title{Astrophysical magnetic field reconstruction\\
and spectroscopy\\
with ultra high energy cosmic rays}

\author{Diego Harari$^a$, Silvia Mollerach$^b$ and  Esteban Roulet$^b$
\\$^a$Departamento de F\'\i sica, FCEyN, Universidad de Buenos Aires
\\Ciudad Universitaria - Pab. 1, 1428, Buenos Aires, Argentina
\\$^b$CONICET and Centro At\'omico Bariloche\\Av. Bustillo km 9.5, 8400,
S.C. de Bariloche, Argentina
\\ Email: \email{harari@df.uba.ar,mollerac@cab.cnea.gov.ar,
roulet@cab.cnea.gov.ar}}

\abstract
{The next generation of ultra high energy cosmic ray experiments will probably
detect several dozens of events clustered around the direction towards each
of the most powerful extragalactic sources. We develop a method which could 
make possible to reconstruct, from the arrival directions and energies of the 
clustered events, the strength and coherence properties of the magnetic field
along the line of sight towards the sources. The method exploits peculiar 
signatures arising from magnetic lensing effects, such as the strong flux 
magnification of multiple images around caustics. We also discuss how to 
obtain information about the cosmic ray composition, and apply this method 
to samples of simulated data.}

\preprint{.}

\begin{document}

\section{Introduction}

The origin and nature of ultra high energy cosmic rays (UHECRs), after several
decades of study, is still puzzling \cite{wa00,ol00}. If they are charged 
particles, such as protons or heavier nuclei, their trajectories are affected 
by the magnetic fields along their path from the sources to the observer. 
For energies below $Z\times 10^{18}~$eV ($Ze$ is the CR electric charge)
they are confined by Galactic magnetic fields. For larger energies the 
galactic field is not strong enough to confine them, which combined with the 
lack of any significant excess from the galactic disk suggests
that UHECRs are most probably of extragalactic origin. However, due to the 
effects of the magnetic fields present along their path, they do not point
to their birth places, complicating the task of identifying their
sources. The small angular scale clustering observed  in the AGASA
data \cite{ta99} already hints to the existence of UHECR point
sources \cite{ti01,ta01,is02}, but the small present statistics
does not allow to solve the source identification question. With the
next generation of detectors, like AUGER \cite{auger} and EUSO \cite{euso},
larger clusters are expected, having each several  dozens of events.
If this were the  case, it would  not only
become possible to reconstruct the source positions, but also the data
could be used to obtain information about the magnetic field along the
CR path \cite{wa96,le97,si98,al01}.

In this paper we discuss some strategies which can be adopted
to reconstruct the main parameters of Galactic magnetic fields from a set 
of events originating from an extragalactic point source, and test them using
simulated data.  The knowledge of the magnetic field
responsible for the deflections allows to better reconstruct the
locations of the sources and, moreover,  
can also be helpful to do spectroscopy of
CRs, making possible the  measurement of  their charges.
The method we develop here profits from several specific features
imprinted on the events by the  magnetic fields,
which not only deflect the trajectories but also
lead to strong lensing phenomena, including the formation of multiple
images and energy dependent magnifications or demagnifications of the
CR fluxes \cite{ha99,ha00,ha02}.

\section{Magnetic lensing}

The magnetic field in the Galaxy is known to have a regular and a
turbulent component \cite{han01}. The regular field in the disk follows
the spiral arms structure, with reversals taking place from arm to
arm. The local value is $B_{reg}\simeq 2~\mu$G. The possible existence 
of an extended halo field (of few kpc scale height) is not settled down yet.
The typical deflection in the direction of propagation of a CR
particle of charge $Ze$ and energy $E$ produced by a regular magnetic field
is
\begin{equation}
\delta\simeq 8.1^\circ \frac{40\ {\rm EeV}}{E/Z}\left|\int_0^L
\frac{{\rm d}{\bf{s}}}{3\ {\rm kpc}}
\times\frac{\bf{B}}{2\ \mu {\rm G}}\right|,
\label{delta}
\end{equation}
where $1~{\rm EeV}=10^{18}$~eV.

The turbulent galactic magnetic field component has a root mean
square amplitude larger than the regular field, 
$B_{rms}\simeq 1$--$3  B_{reg}$, and the largest
turbulence scale is $L_{max}\simeq 100$ pc, while the minimum one is
believed to be much smaller ($L_{min}\ll L_{max}$).
It is often modeled as a Gaussian random field with a Kolmogorov 
spectrum \cite{ar81} (i.e. such that the energy density satisfies 
d$E/{\rm d}k\propto k^{-5/3}$).
The mean deflection of particles moving in the turbulent field
vanishes, while the root mean square deflection is given by
\cite{ha02}
\begin{equation}
\delta_{rms}\simeq 1.4^\circ \frac{40~ {\rm EeV}}{E/Z}
\frac{B_{rms}}{4~ \mu {\rm G}}
\sqrt{\frac{L}{3\ {\rm kpc}}}\sqrt{\frac{L_c}{50\ {\rm pc}}},
\label{deltarms}
\end{equation}
where $L_c$ is the coherence length. For a narrow band ($L_{min}
\simeq L_{max}$) spectrum, one has $L_c\simeq L_{max}/2$, while for a broad
band ($L_{max}\gg L_{min}$) Kolmogorov spectrum it is $L_c\simeq
L_{max}/5$.

We have written the random and regular field deflections
in terms of the characteristic values of the Galactic fields, thus the
numerical value of the deflections are typical expected values. We see
that the regular field generally produces the dominant effect on the
deflection, which leads to a coherent displacement of the apparent 
position of the source that is inversely proportional to the energy. 
Compared to this, the random field produces rather small changes in the 
arrival direction of the CRs. However, as the charged particles propagate
through the magnetic field, they are not only deflected, but their flux
is also focused or defocused due to the differential deflections of
neighboring paths \cite{ha99,ha00}, with the effects being
larger at smaller energies.
As lower energies are considered, CR particles from one source
can have more than one path leading to the detector, each traversing an
uncorrelated magnetic field patch, and this leads to the appearance of
multiple images of the source. This typically occurs when 
$\delta_{rms}\simeq L_c/L$~\cite{wa96,ha02}, what allows to define a 
critical energy $E_c$, around and below which the formation of multiple 
images is very likely, through\footnote{
The regular field can also lead to multiple images,
with its characteristic scale of homogeneity (few kpc) playing the role
of the coherence length.}
\begin{equation}
\delta_{rms}\equiv \frac{E_c}{E}\frac{L_c}{L}.
\label{ecdelta}
\end{equation}
Its numerical value is given by\footnote{
If the amplitude of the turbulent field were not
constant, but had instead some modulation along the Galaxy, the factor
$B_{rms}\sqrt{L}$ in eq.~(\ref{deltarms}) should be replaced by
$|\int_0^L {\rm d}s~B^2_{rms}(s)|^{1/2}$, and
$B_{rms}L^{3/2}$ in eq.~(\ref{Ec}) should be replaced
by$\left[3\int_0^L{\rm d}x\
x\int_0^x{\rm d}s\ B^2_{rms}(s)\right]^{1/2}$.}
\begin{equation}
E_c\simeq Z\ 60~ {\rm  EeV}\frac{B_{rms}}{4\ \mu{\rm G}}
\left(\frac{L}{3\ {\rm kpc}}\right)^{3/2}\sqrt{\frac{\rm 50 \ pc}{L_c}}.
\label{Ec}
\end{equation}
The new images appear in pairs, at an angular distance $\sim L_c/L$ from 
the position of the original source image and  with a large magnification 
of their fluxes. The peak in the amplification when new images
appear leads to small angular scale clustering and to an excess of events
in an energy bin close to $E_c$ \cite{ha00,ha02}.

The overall picture can be summarized as follows: the magnetic fields
present along the CR trajectories change the mapping between the
observed arrival directions and the source ones. At the highest energies,
the effects are small and hence the mapping is close to the identity. 
Going down in energy, deflections arise due to the turbulent and the regular
magnetic field components. The turbulent component leads to small {\it
rms} deflections, but it gives rise to a network of lines in the sky
around which the fluxes of potential sources would be significantly
magnified. This network corresponds to the directions where folds in the 
mapping (caustics) develop. The caustics appear at energies near (and below) 
$2~E_c$, giving rise to multiple images of
the sources lying close to those directions \cite{ha02}.
The regular field, which is responsible for larger
deflections, leads to a global displacement of this network of
caustics relative to the source directions. As a result of this
drift caused by the regular field, a given CR source will experience 
successive magnifications and demagnifications of its flux at different 
energies caused by the motion of the caustic network across the sky. 
For decreasing energies, the continuous appearance of new images also leads 
to additional peaks in the spectrum, and strong lensing effects
associated to the regular field component will also show-up.

\section{Magnetic field reconstruction}

The properties of the magnetic deflections and lensing effects can be used 
to determine the magnetic field parameters from the study of
clustered UHECR events having their  origin in
isolated point-like sources. For this purpose one should start by
defining  the regions containing the excess events
(which have to be larger than at least a few times the angular
resolution of the detector but yet not too large to avoid excessive
background events coming from other independent nearby sources).
Different strategies must be considered, depending on
the relative strength of the effects due to regular and random
components, and in what follows we develop several ideas to
perform this reconstruction.

Consider first the likely case in which the deflection caused by a regular
component (either galactic or extragalactic) is larger than the
rms deflection imprinted  by the turbulent galactic field.
In this case,  the overall angular motion of the source images as a function
of energy can be used to reconstruct the integral along the line of sight to 
the source of the perpendicular component of the regular magnetic field.
In order to do this, one first finds the overall direction of motion of the 
images caused by the regular field by fitting a straight line  
($\alpha_y=a+b \alpha_x$) to the event coordinates  $\alpha_{x,y}$.
The integral along the line of sight of the
perpendicular component of the regular magnetic field will just be at a right
angle with respect to this line, and its absolute value can be
obtained choosing now the coordinates $\alpha_\parallel$
(in the direction along the deflections produced by the
regular field) and $\alpha_\perp$ (orthogonal to it), and plotting 
$\alpha_\parallel$ vs. $1/E$ for the set of selected events.
One should have, ignoring at this step the random field component, that
\begin{equation}
{\alpha_\parallel}_i \simeq \alpha_0+\frac{Z_i}{E_i} K_\perp
\label{alpar}
\end{equation}
with
\begin{equation}
K_\perp \equiv \pm |e\int_0^L {\bf {\rm d}s}\times{\bf B}^{reg}|\simeq
\pm 1.3^\circ (40~{\rm EeV})\left|
\int_0^L\frac{\bf {\rm d}s}{\rm kpc}\times \frac{{\bf B}^{reg}}{\mu
{\rm G}}\right|.
\end{equation}

If the CRs have all identical electric charge $Z$, the events  would
fit nicely to just one straight line, $\alpha_\parallel=\alpha_0+c/E$. 
The slope $c$ would give the value of $Z~K_\perp$, which essentially provides
the  magnitude and direction of the
regular magnetic field orthogonal to the l.o.s. multiplied by its
relevant scale length, modulo the CR electric charge. The value of
$\alpha_0$ just gives the
original location of the source ($\alpha_\parallel=\alpha_0$,
$\alpha_\perp=0$). After the source location is estimated, it proves
convenient to change variables to $\alpha_\parallel\to {\rm sign}
(K_\perp)\,(\alpha_\parallel-\alpha_0)$, defined such that it
tends to zero at large energies and increases as $E$ decreases. In
what follows we will use the name  $\alpha_\parallel$ just to refer to
this shifted variable.

We illustrate this procedure using simulated data, obtained through a
ray shooting technique, in which a large number of antiparticles were
thrown isotropically from Earth, their trajectories numerically
integrated across a distance $L$ within a homogeneous magnetic field plus 
a turbulent component, and were finally recorded only if they pointed
within a small cone (of aperture $\sim \delta_{rms}/10$)
from the direction to a fixed `source'.
The source was assumed to inject CRs with a differential
energy spectrum ${\rm d}N/{\rm d}E\propto E^{-2.7}$.
About 100 simulated events with energy above $40$ EeV  were selected, 
assuming the detector efficiency to be energy-independent.
We display the `theoretical' results from the simulation (with
black dots) as well as more realistic ones (open circles) obtained by
adding noise to the energy (corresponding to a detector energy
resolution of 10\%) and to the angles (corresponding to
an angular resolution of $0.2^\circ$ in each direction).

In the case illustrated in Figure~\ref{mono} the homogeneous magnetic 
field had strength $2\ \mu$G along the $\alpha_x$ direction, the turbulent 
component had  $B_{rms}=4\ \mu$G with just one turbulence scale $L_{max}
=L_{min}=100~$~pc (and thus $L_c=50~$pc), and the distance traversed by the
CRs (assumed to be protons) within these fields was $L=3~$kpc.
The source actual position was in the direction ($\alpha_x=0,\alpha_y=0$).
The first panel displays the simulated data, with the straight line
being the reconstructed overall direction of
motion of the images and the star being the reconstructed source location.
The second panel plots instead $\alpha_\parallel$ vs. $1/E$, while the
third one will be discussed farther below.

\FIGURE{
\epsfig{file=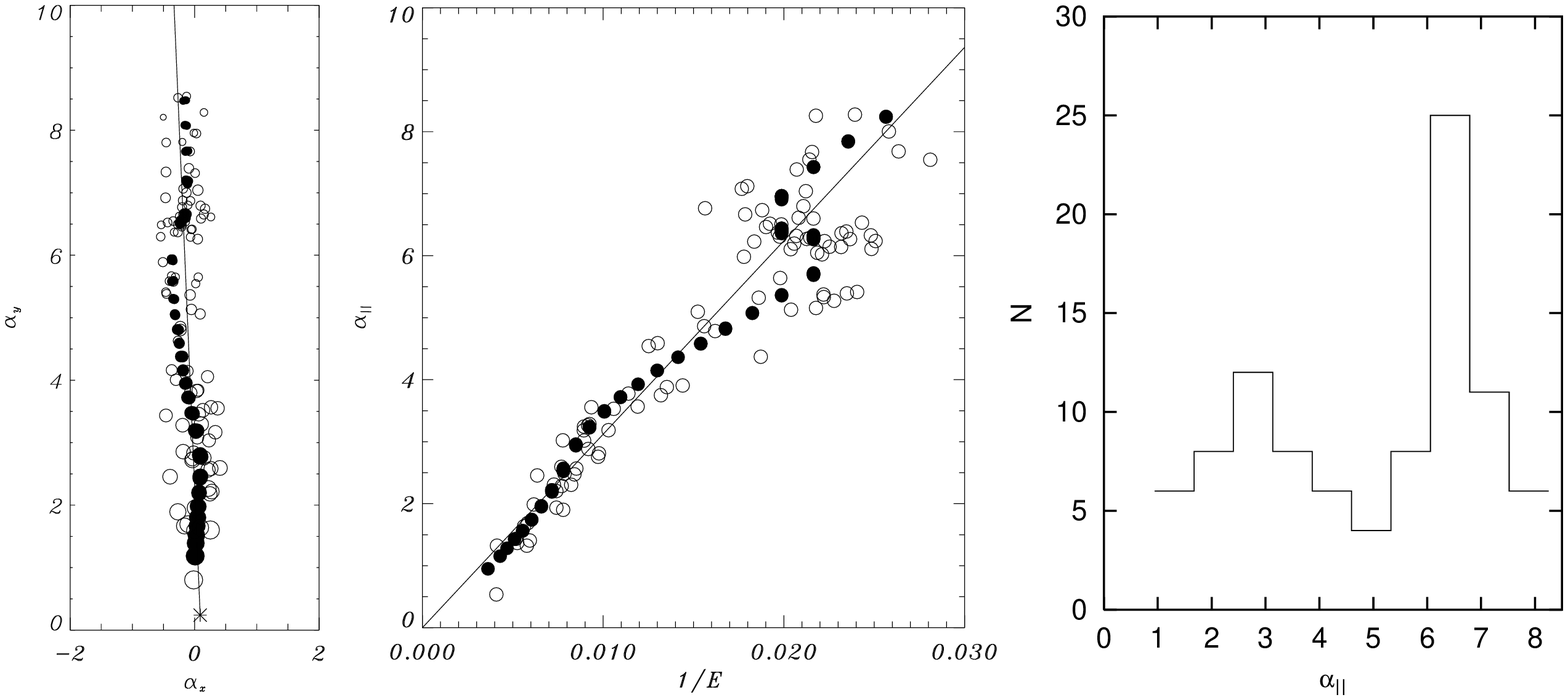,width=15truecm}
\caption{First Panel: Angular distribution of the simulated events for
a regular field $B_{reg}=2\ \mu$G, a turbulent field with
$B_{rms}=4\ \mu$G, $L_{max}=L_{min}=100$~pc and $L=3$~kpc. White circles
have a Gaussian noise with dispersion $0.2^\circ$ added in each
direction, as well  as a 10 \% uncertainty in the energy,
 to simulate the detector sensitivity, while black circles
have no noise added. The size of the circle grows with the energy of
the event. The asterisk indicates the reconstructed position of the
source, and the solid line the fit to the deflection due to the regular field.
Second panel: $\alpha_\parallel$ vs. $1/E$~[EeV$^{-1}$] and the linear 
fit to the data. Third panel: Number of events vs. $\alpha_\parallel$ 
divided in ten angular bins.}
\label{mono}}

In Figure~\ref{kol} we display the results for another simulation with
the same parameters for the regular field and same root mean square
amplitude of the turbulent field but with a Kolmogorov spectrum with
$L_{max}=10 L_{min}= 100$~pc (corresponding to $L_c=25$~pc).
The result for the best fit of the numerical reconstruction was for
both cases $Z|\int_0^L{\rm d}{\bf s}\times {\bf B}|\approx 6.1~ \mu$G~kpc, 
in good agreement with the values used in the simulations.

In the case of a mixed composition, the fit of $\alpha_\parallel$ vs. $1/E$
to just one straight
line should fail (excessive $\chi^2$), since one expects to see the
superposition of several straight lines, all with the same value of
$\alpha_0$ but different slopes, with ratios fixed by the ratio of the
different electric charges $Z_i$ involved\footnote{The value of
$\alpha_0$ in the case of mixed composition was just determined by eye,
looking for the value of $\alpha_\parallel$ towards which the different
lines converged.}. In this case one can separate
the events in groups, above and beyond limiting slopes,
and then fit each group independently to an expression of the form
given in Eq.~(\ref{alpar}).
This allows to obtain information on both $K_\perp$ and the ratio of
charges $Z_i$ involved, besides the relative amounts of different
nuclei from the relative number of events on the different lines.
Notice that the reconstructed fraction of different nuclei
reflects the composition upon arrival to Earth, which may differ from
the fraction injected by the source within the same energy range
because the flux magnification due to magnetic lensing depends on
the combination $E/Z$. Thus, the flux of different
nuclei is magnified by a different factor at a given energy \cite{ha00}.
\FIGURE{
\epsfig{file=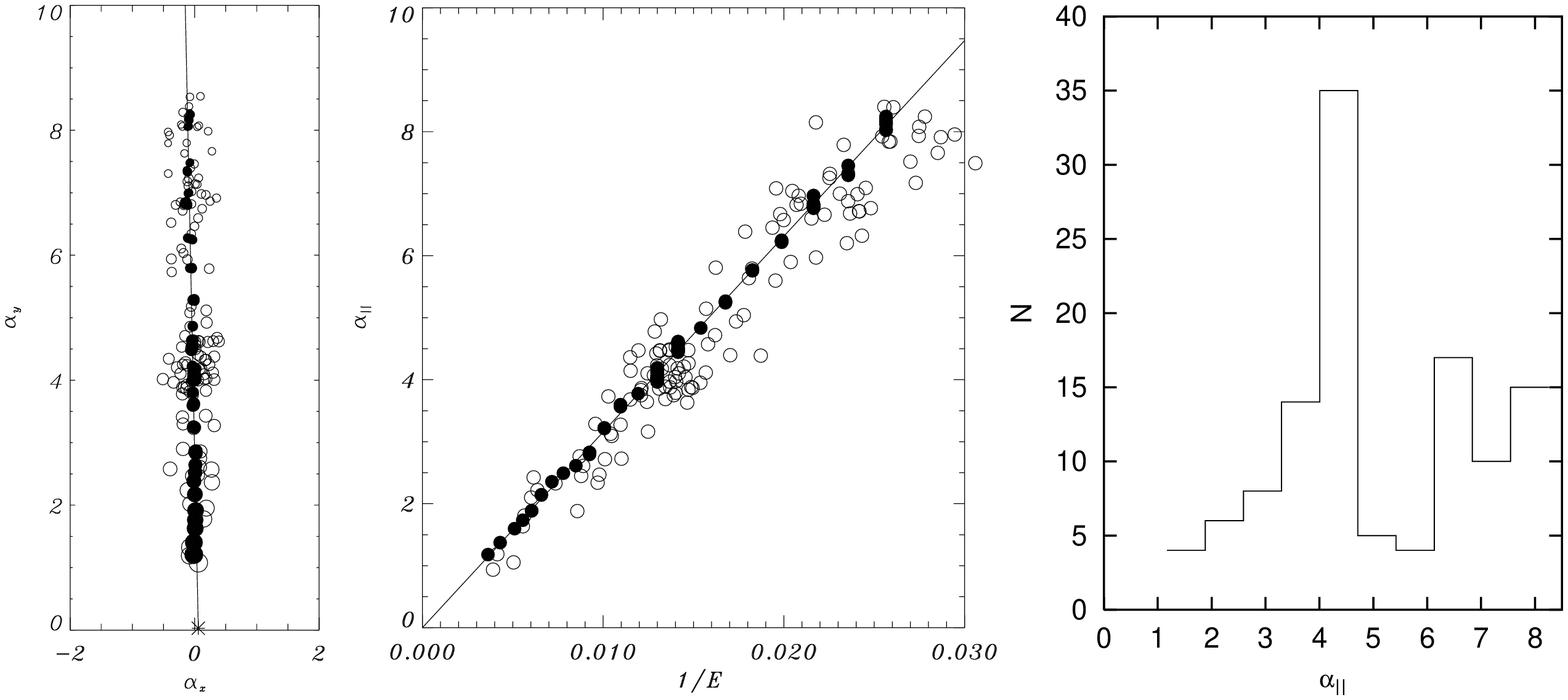,width=15truecm}
\caption{Same as Figure~\ref{mono} for a turbulent field with
a Kolmogorov spectrum and $L_{max}=10 L_{min}=100$~pc.}
\label{kol}}

In Figure~\ref{kolz3} we exemplify this procedure for the same values
of the fields and Kolmogorov spectrum with 60\% of protons and 40\% of
Lithium ($Z=3$). (The actual fraction of Lithium injected by
the source was 50\% in this case.)
The two slopes in $\alpha_\parallel$ vs. $1/E$ are
clearly separated and the reconstructed values are
$Z_1|\int_0^L{\rm d}{\bf s}\times {\bf B}| \approx 5.9~\mu$G~kpc,
and  $Z_2|\int_0^L{\rm d}{\bf s}\times {\bf B}| \approx 18.6~\mu$G~kpc,
implying that $Z_2/Z_1\approx 3.1$.

In Figure~\ref{kolz2} we show the analogous results for 80\% of protons
and 20\% of Helium, for which we obtain
$Z_1|\int_0^L{\rm d}{\bf s}\times {\bf B}| \approx 6.2~\mu$G~kpc,
and  $Z_2|\int_0^L{\rm d}{\bf s}\times {\bf B}| \approx 12.3~\mu$G~kpc,
implying that $Z_2/Z_1\approx 2$.  The reconstructed
direction of the regular magnetic field was off by only a few degrees
in all cases. Let us mention that we performed several other
simulations with different realizations of the random field component,
obtaining reconstructed values of similar quality.

\FIGURE{
\epsfig{file=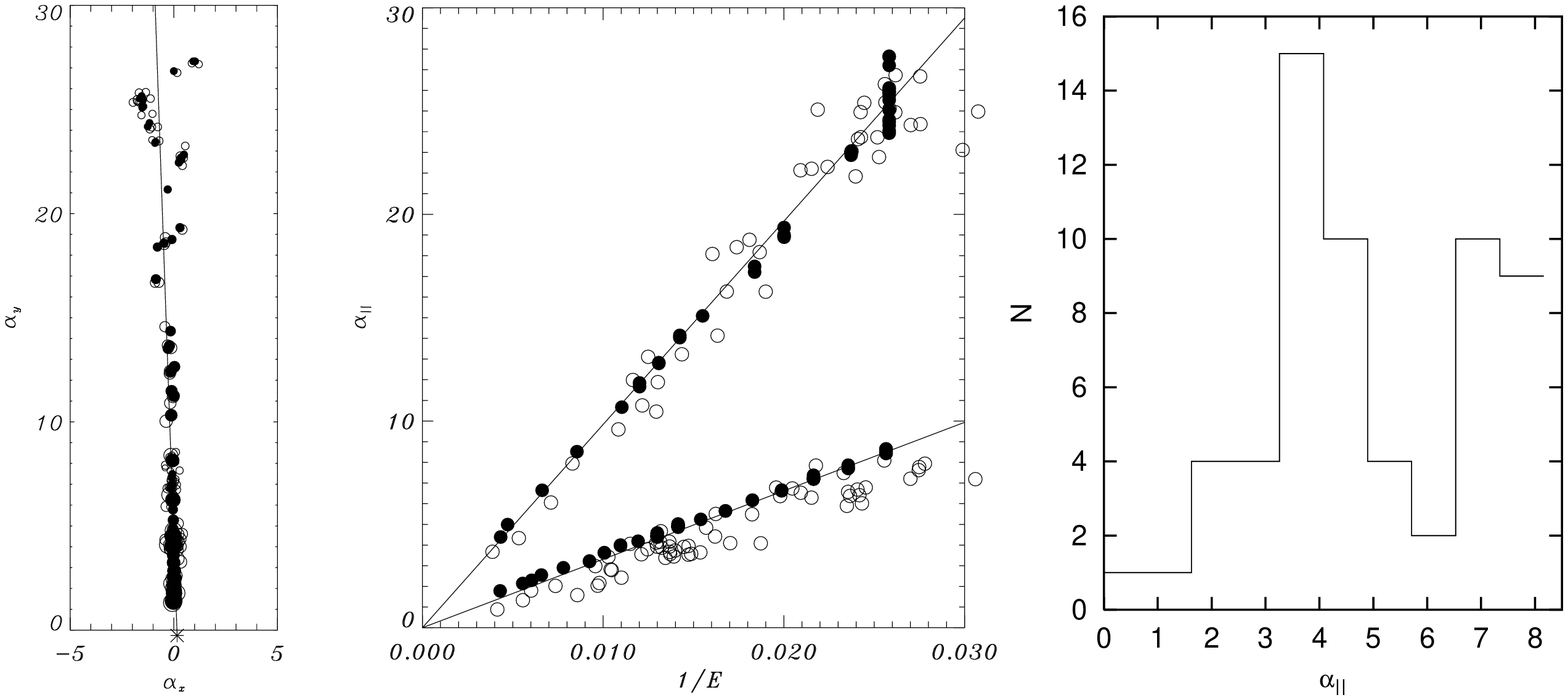,width=15truecm}
\caption{Same as Figure~\ref{kol} with mixed composition, 60\% of
protons and 40\% of Lithium.}
\label{kolz3}}

\FIGURE{
\epsfig{file=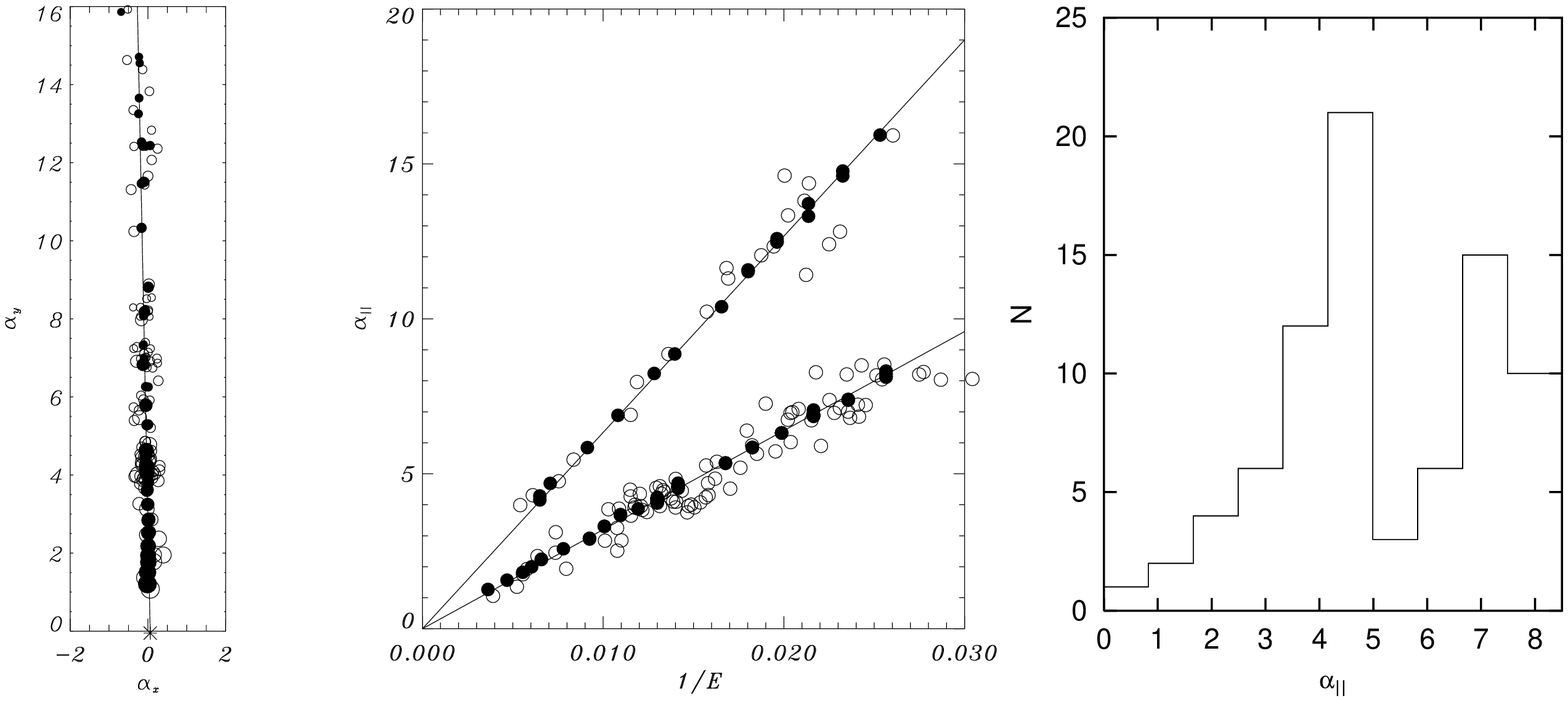,width=15truecm}
\caption{Same as Figure~\ref{kol} with mixed composition, 80\% of
protons and 20\% of Helium.}
\label{kolz2}}

Having determined the overall motion of the images due to the regular
magnetic field component, we turn to analyze the effects of the
turbulent field. At high energies the main effect is the formation of
multiple images, with their associated peaks manifesting in the
observed flux. The critical energy of this process, $E_c$ (given by 
eq.~(\ref{Ec})), can be determined from the location of the first lensing 
peak, which is expected to lie at energies somewhere between $2E_c$ and 
$E_c/2$. We then take bins in
$\alpha_\parallel$ covering the range of angles where clustered events
are observed above a certain energy threshold. The bins should be
larger than the angular resolution of the detector and such that a
sizeable number of events per bin result (taking e.g. ten angular bins
if the number of clustered events is of the order of one hundred). Now
one can plot the number of events in each bin
vs. $\alpha_\parallel$. If there were only deflections but no flux
(de)magnification due to lensing effects, the regular
field would just lead to $\Delta
N/\Delta\alpha_\parallel\propto \alpha_\parallel^\beta$, where $\beta$
characterizes the slope of the CR energy spectrum, with ${\rm d}N/{\rm
d}E\propto E^{-2-\beta}$.

The advantage of plotting $\Delta N$ in terms of $\alpha_\parallel$
(and not e.g. vs. $E^{-1}$) is that regardless of the CR composition
the first lensing peak always appears at the same value of 
$\alpha_\parallel$. Referring to this value as $\alpha_\parallel^{peak}$, 
one has then (up to an overall factor of two)
\begin{equation}
\frac{E_c}{Z}\simeq \frac{|K_\perp|}{\alpha_\parallel^{peak}}
\end{equation}

The third panels of Figures~\ref{mono}-\ref{kolz2}  illustrate the
results of this method applied to the simulated data.
In Figure~\ref{mono} the first lensing peak associated to multiple
image formation appears
at $\alpha_\parallel^{peak}\simeq 6.4^\circ$ and corresponds to the
sharp peak. It leads to a critical energy value
$E_c/Z \approx 49~$EeV,
in good agreement with the parameters used in the simulation (which
correspond to $E_c\approx 60$~EeV). A shallower and more symmetric
excess of events is also evident at smaller deflection angles,
$\alpha_\parallel^{peak}\simeq 2.5^\circ$, and it is due to a high
magnification region crossing the source location at high energies
(larger than $\approx 2 E_c$, when folds were not yet formed).

In Figure~\ref{kol} the first peak appears at
$\alpha_\parallel^{peak}\simeq 4.3^\circ$, corresponding to
$E_c/Z\approx72$~EeV (while the simulation parameters lead to
$E_c/Z\approx 85$~EeV). For the mixed composition cases
(Figures~\ref{kolz3} and \ref{kolz2}) the corresponding values are
$E_c/Z\approx 85$~EeV and $E_c/Z\approx 87$~EeV respectively.

The angular distribution of the events can also be used to estimate
$L_c/L$ ($L$ here is the distance traversed along the turbulent component, 
which needs not be the same as that across the regular one).
There are two independent ways to do this.
One is based on the fact that at energies $\sim~E_c$
the typical angular separation in the network of caustics of the
turbulent magnetic field is $4L_c/L$ \cite{ha02}. Since the effect of
the regular field  induced deflections is to move these caustics,
relative to the source direction, the typical
separation in $\alpha_\parallel$ between consecutive lensing peaks
is also equal to $4L_c/L$.
If at least two lensing peaks are observed in the data from one cluster,
then $L_c/L$ can be estimated. This is the case in the simulations
presented above. In the case with just one wavelength depicted in
Figure~\ref{mono} (which has $L_c=50$~pc), the separation between
peaks is of order $4^\circ$, as it can be seen in the third panel,
leading to $L_c/L\approx 0.017$. In the Kolmogorov spectrum cases (for
which $L_c\simeq 25$~pc) depicted in Figures~\ref{kol}, \ref{kolz3} and
\ref{kolz2}, the separation among peaks is between $2^\circ$ and $3^\circ$,
leading to $L_c/L$ values in the range 0.008--0.013. All these values
are within 50\% of the actual values used in the simulations.

Notice that it is likely to observe the first lensing peak rather
close to $2E_c$, and to observe more than one lensing
peak in the range of energies between $2E_c$ and $E_c/2$, if the deflection
caused by the regular magnetic field component is strong enough.
Indeed, the network of caustics is already well defined at energies around 
$2E_c$. Although only a small percentage
of all potential source directions are sufficiently close to a caustic
at energies around $2E_c$ to have their flux noticeably magnified,
if the deflection caused by the regular magnetic field component
is significant, a caustic will reach the source position at energies
not far below $2E_c$. For instance, if the regular field deflection is
at least 3 times larger than the rms deflection caused
by the random field, then the source apparent position is displaced,
as the energy decreases from $2 E_c$ down to $E_c/2$, by more than
$4L_c/L$. Since this is the typical separation between caustics, it is
likely to produce more than one lensing peak in this energy range.
Notice however that if the deflection by the regular field were too strong, 
the caustics may be crossed too fast, leading to a small integrated effect 
and thus a less noticeable peak.

Alternatively, $L_c/L$  can be determined
from the fact that it is just the typical angular scale
characterizing the separation between the images present in the first
lensing peak. In this peak there are in principle one image
corresponding to the principal image which is also present at higher
energies, and two new images which are both in the same location of
the sky and have appeared separated from the previous one by an
angular scale of order $L_c/L$. Since this angular separation
may be quite small, this
approach can be used only when the angular uncertainty in the events
is smaller than this scale. For the example shown in
Figure~\ref{mono} this separation is  about $1^\circ$,
leading to an $L_c/L$ estimate similar to the one obtained with the
previous method. For the Kolmogorov spectrum cases, that leads to
smaller $L_c/L$, we see from the first panel in Figures~\ref{kol},
\ref{kolz3} and \ref{kolz2} that to apply this
method in cases like these would require a quite precise angular
resolution, probably better than the one achievable with AUGER ($\sim
0.3^\circ$), but possibly within the aim of some proposed detectors
\cite{sa02}.

Once $E_c/Z$ and $L_c/L$ have been estimated, we can combine them to
obtain an estimate of $B_{rms}~L$ using Eq.~(\ref{Ec}). In our
examples, the recovered values fall in the range
$B_{rms}~L\simeq 10$--$13~ \mu$G~kpc, in good
agreement with the value $12~ \mu$G~kpc used in the simulations.

If the deflection caused by the regular magnetic field is 
smaller than the rms deflection caused by the turbulent field at the
same energy, the overall motion of the images does not provide reliable
information about the regular magnetic field component, other than an
upper bound on its integrated effect. 
However, it might still be possible to better reconstruct the source
location by looking at the highest energy events, i.e. at energies 
above those at which the multiple images appear. Indeed, as long as the 
deflections are small ($\delta\ll L_c/L$), the random field component still
produces a coherent deflection of the images proportional to
$(E/Z)^{-1}$, and hence unless there is a significant mixture of
different CR compositions there will be a clear signal allowing to
reconstruct the source location.  One could
also extract information about the turbulent field through the
spectral features imprinted by the lensing peaks, and by the angular
clustering properties of the events. The critical energy 
$E_c$ can be estimated as the highest energy at which a peak is observed 
in the spectrum. With sufficient angular resolution, $L_c/L$ can be 
estimated from the angular separation between the images present in the 
first lensing peak, while information about the composition can be extracted 
from the observed energy ratios of the events clustered in the peaks
\cite{ha02}. 
\FIGURE{
\epsfig{file=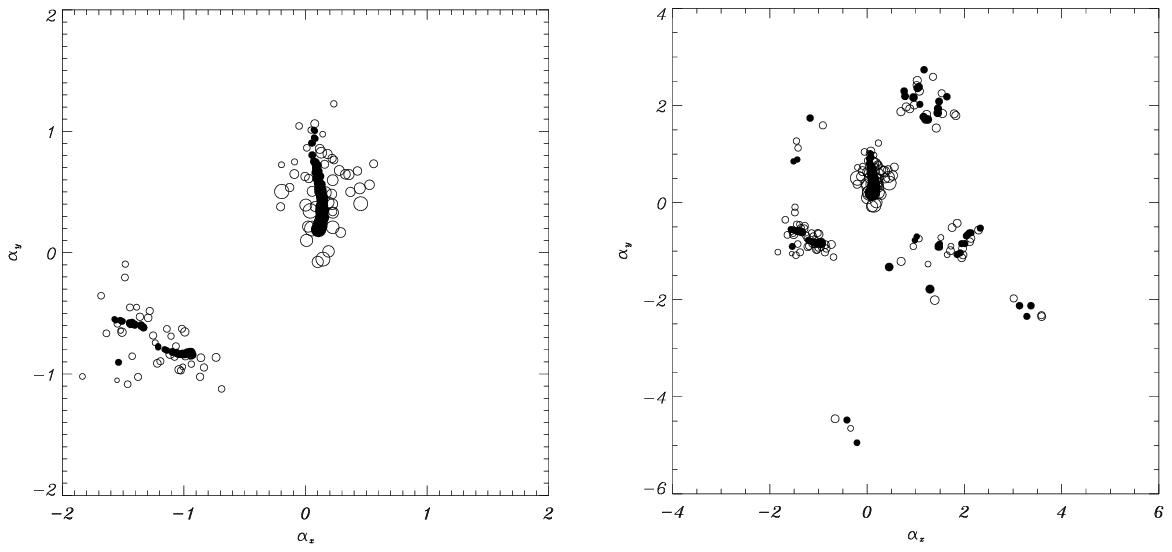,width=15 truecm}
\caption{Simulated data in a regular field $B_{reg}=0.1\ \mu$G,
and other parameters as in Fig.~\ref{mono}. The injected composition is
100\% protons (left panel), and a mixture with 40\% Beryllium (right
panel).}
\label{sp}}

These general features are illustrated in Figure~\ref{sp}, which
corresponds to a ray shooting simulation with a small orthogonal
component of the regular magnetic field (leading to $\int
B_{\perp}^{reg}{\rm d}s=0.3\ \mu$G~kpc). The random component has
$B_{rms}=4\ \mu$G (and spectrum with $L_{min}=L_{max}=100$~pc) and
 $L=3$~kpc. Events with energies larger than 40~EeV and
injection spectrum $\propto E^{-2.7}$ are displayed. The left panel is
for CR protons alone. The highest energy events move away from the
source location (here the origin) with a deflection $\propto E^{-1}$,
and at energies near 60 EeV two pairs of secondary images appear. They
are displaced by $\sim 1.5^\circ$ from the principal image, hence
leading to an estimated $L_c/L\simeq 0.026$, in good agreement with
the input value 0.017.  For the right panel, which has a mixed
composition of protons and 40\% Be at injection, other secondary
images appear for $E>40$~EeV, dispersed over a region of $\sim
3^\circ$, and they contain exclusively the heavier Be nuclei, since the
protons in them have energies below 40~EeV. In general each of the
secondary images has a narrow rigidity distribution due to the lensing
origin of the peaks (and hence if two different elements were
contributing to an image in the considered energy range, the events
would split into groups with energy ratios given by the ratios of
their charges).

\section{Conclusions}

One of the most exciting perspectives for the Auger Observatory currently
under construction, as well as for other future experiments, is that they may
inaugurate the era of UHECR astronomy. A hint that this may be the case is
the small angular scale clustering observed by AGASA, which may be the
first indication that a few UHECR sources are significantly more
powerful than the otherwise apparently isotropic background.

UHECR astronomy is inevitably tied to the magnetic fields along the line
of sight towards the most powerful sources, if the bulk of their emission
is in the form of charged particles, such as protons or heavier nuclei.
Conservative estimates of galactic and intergalactic magnetic fields indicate
that the deflection of protons should be small at the highest energies,
around and above $10^{20}$~eV, so that they should truly point to their
birthplace. However, even protons with energies around $10^{20}$~eV may have
their flux significantly (de)magnified by lensing effects, for instance by
the turbulent component of the magnetic field in the Milky Way. A heavier
component of UHECRs would undergo deflections larger than protons at
comparable energies, and would be subject to similar flux (de)magnification
effects at higher energies.

Consequently, the behaviour of the transition towards UHECR astronomy at the
highest energies is a very rich source of information both on the source
properties as well as on the characteristics of the intervening magnetic
fields. This has been exploited, for instance in \cite{si98}, as a tool for
the reconstruction of intergalactic magnetic field parameters, particularly
in the case of bursting sources of UHECRs \cite{wa96}, where the
energy-dependent time delays provide additional handles to perform the task.

Here we developed a strategy to reconstruct the parameters of intervening
magnetic fields that profits from the strong lensing effects that take place
at energies of the order of $E_c$, given by Eq.~(\ref{Ec}), around which it is
likely to observe strongly magnified multiple images of single UHECR sources.
We have shown that it is likely that a sort of `magnetic spectroscopy' of 
UHECR sources may allow to measure the strength of the magnetic field along 
the line of sight, its coherence properties, and the relative amounts of 
nuclei with different electric charges among the observed CRs.
Our analysis assumed the sources to be steady, what means here
essentially that the typical time delays between different images are
smaller than the  timescales of  emission of the  sources.

The method works if the number, strength and location of sources, the UHECR 
composition, and the intervening magnetic field parameters are such that 
clusters with a large number (several dozens) of events can be identified, 
without excessive background from events with origin in nearby independent 
sources. The angular and energy resolution required is also dependent on the 
actual values of the parameters involved. We have shown, through illustrative 
examples based in simulated data, that the method is likely to be applicable 
for realistic values of the parameters of the galactic magnetic field. 
The overall angular 
motion of the events as a function of energy allows to reconstruct the 
integral along the line of sight of the perpendicular component of the 
regular magnetic field, modulo the CR electric charge. If this can be
done for several sources, one may then even map the strength and
extent of the magnetic field along different directions, and this
could help to establish its overall distribution.\footnote{Notice that
the information reconstructed with UHECRs is complementary to that
provided by Faraday rotation measurements, which are sensitive to the 
magnetic field component parallel to the line of sight.} 
The reconstructed magnetic field also provides a 
handle to measure the relative number of CRs with different electric charges. 
A `spectroscopic' analysis of the peaks in the number of events as a 
function of this overall motion (or as a function of energy when the 
effect of the regular component is not strong enough) provides a measure 
of the critical energy $E_c/Z$ at which strong lensing phenomena occur. 
The enhanced angular clustering of the events at the energies where
strong flux magnification takes place can also be used to estimate $L_c/L$, 
the ratio between the coherence length of the turbulent magnetic field 
component and the path length traversed by the CRs across it. 
$E_c/Z$ and $L_c/L$ provide a measure of $B_{rms}L$, independent of the 
CR electric charge. The fact that the magnification peaks
due to magnetic lensing occur at energies which are in direct proportion to 
the electric charge $Z$ provides a second handle to grasp the CR composition. 
It induces a strong correlation of arrival directions at energy ratios fixed 
by the ratios of the CR electric charges.

Potentially, this method could also extract information about the different 
scales involved in the turbulence of the intervening magnetic field, if the 
experiments had sufficient energy and angular resolution (and, of
course, statistics). If the spectrum is 
broad band, long and short wavelengths lead to amplification peaks at 
somewhat different values of $E/Z$~\cite{ha02}. The density of lensing peaks 
could thus be used to obtain information on the magnetic field spectral 
properties. However, in realistic experiments the very narrow lensing peaks 
due to short wavelengths are likely to go unnoticed.

The method may also provide information on extragalactic magnetic fields.
Notice that for an extragalactic turbulent component one has an
associated critical energy
\begin{equation}
E_c\simeq Z\ 2\times 10^{19}~{\rm EeV} 
\frac{B_{rms}}{10^{-9}~ {\rm G}}
\left(\frac{L}{10\ {\rm Mpc}}\right)^{3/2}\sqrt{\frac{\rm Mpc}{L_c}}.
\label{ecext}
\end{equation}
Hence, if these magnetic fields have a large extent ($L\gg {\rm Mpc}$)
and a significant strength ($\gg 10^{-9}$~G), their effects could
produce noticeable signals at energies above $10^{20}$~eV even for
protons, while for more modest fields, it would be the galactic magnetic 
fields the ones producing the dominant effects.

\acknowledgments

Work partially supported by ANPCYT, CONICET, and Fundaci\'on
Antorchas, Argentina.


\begin{thebibliography}{100}

\bibitem{wa00}
A.A. Watson, \emph{Ultrahigh-energy cosmic rays:
The experimental situation}, \prep{333}{2000}{309}.

\bibitem{ol00}
A. Olinto, \emph{Ultrahigh-energy cosmic rays:
The theoretical challenge}, \prep{333}{2000}{329}.

\bibitem{ta99}
N.~Hayashida et al., \emph{Updated AGASA event list above
$4\times 10^{19}\,eV$}, \apj{522}{1999}{225}  [\astroph{0008102}].

\bibitem{ti01}
P.G. Tinyakov and I.I. Tkachev, \emph{Correlation function of ultrahigh energy
cosmic rays favors point sources},  \jetpl{74}{2001}{1} [\astroph{0102101}].

\bibitem{ta01}
M. Takeda et al., \emph{Clusters of cosmic rays above $10^{19}$ eV
observed by AGASA}, in \emph{Proc. ICRC} Hamburg, Germany 2001, p.~341;\\
M. Teshima, talk at TAUP 2001, Gran Sasso, Italy (2001), in press.

\bibitem{is02}
C.~Isola and G.~Sigl, \emph{Large Scale Magnetic Fields
and the Number of Cosmic Ray Sources above $10^{19}$ eV}
[\astroph{0203273}].

\bibitem{auger}
http://www.auger.org/

\bibitem{euso}
http://euso-mission.org/

\bibitem{wa96}
E. Waxman and J. Miralda-Escud\'e,
\emph{Images of bursting sources of high-energy cosmic rays,
1. Effects of magnetic fields}, \apj{472}{1996}{L89}
[\astroph{9607059}].

\bibitem{le97}M.~Lemoine, G.~Sigl, A.V.~Olinto and D.N.~Schramm,
\emph{Ultra-high energy cosmic ray sources and large scale magnetic fields},
\apj{486}{1997}{L115} [\astroph{9704203}].

\bibitem{si98}
G.~Sigl and M.~Lemoine, \emph{Reconstruction of Source and
Cosmic Magnetic Field Characteristics from Clusters of
Ultra-High Energy Cosmic Rays}, \app{9}{1998}{65}
[\astroph{9711060}].

\bibitem{al01}
J. Alvarez-Mu\~niz, R. Engel and T. Stanev, \emph{UHECR propagation in
the Galaxy: clustering versus isotropy} [\astroph{0112227}].

\bibitem{ha99}
D. Harari, S. Mollerach and E. Roulet, \emph{The toes of the
ultrahigh-energy cosmic ray spectrum}, \jhep{08}{1999}{022}
[\astroph{9906309}].

\bibitem{ha00}
D.~Harari, S.~Mollerach and E.~Roulet, \emph{Signatures of galactic
  magnetic lensing upon ultra high energy cosmic rays},
  \jhep{02}{2000}{035} [\astroph{0001084}].

\bibitem{ha02}
D.~Harari, S.~Mollerach, E.~Roulet and F. S\'anchez,
\emph{Lensing of ultra high energy cosmic rays in turbulent
magnetic fields}, \jhep{03}{2002}{045} [\astroph{0202362}]

\bibitem{han01}
J. L. Han, \emph{Magnetic Fields in our Galaxy: How Much Do We Know?},
{\it Astrophys.\ Space\ Sc.\ }{\bf 278} (2001) 181
[\astroph{0010537}].

\bibitem{ar81}
J.W. Armstrong, J.M. Cordes and B.J. Rickett,
\emph{Density power spectrum in the local interstellar
medium}, \nature{291}{1981}{561}.

\bibitem{sa02}
M. Sasaki, A. Kusaka and Y. Asaoka,
\emph{Design of UHECR telescope with 1 arcmin resolution and 50 degree
field of view}
[\astroph{0203348}].

\end{thebibliography}
\end{document}